\documentclass[journal,comsoc]{IEEEtran}

\usepackage[T1]{fontenc}

\usepackage{flushend}

\usepackage{cite}

\usepackage{textcmds}

\usepackage{blindtext}

\usepackage{xcolor}
\usepackage{hyperref}
\usepackage[all]{hypcap}
\hypersetup{colorlinks=true, pdfstartview=FitV, linkcolor=blue, citecolor=blue, plainpages=false, urlcolor=blue}

\ifCLASSINFOpdf
   \usepackage[pdftex]{graphicx}
\else
\fi

\usepackage{amsmath}

\usepackage[cmintegrals]{newtxmath}

\begin{document}

\title{Blockchain-based Security Framework for Critical Industry 4.0 Cyber-physical System}
%
%
%


\author{Ziaur~Rahman,
        Ibrahim~Khalil,
        Xun~Yi,
        and~Mohammed~Atiquzzaman,~\IEEEmembership{Senior Member,~IEEE}


}


\markboth{}
{Rahman \MakeLowercase{\textit{et al.}}: IEEE Communications Society Magazine}

\maketitle

\begin{abstract}

There has been an intense concern for security alternatives because of the recent rise of cyber attacks, mainly targeting critical systems such as industry, medical, or energy ecosystem.  Though the latest industry infrastructures largely depend on AI-driven maintenance, the prediction based on corrupted data undoubtedly results in loss of life and capital. Admittedly, an inadequate data-protection mechanism can readily challenge the security and reliability of the network. The shortcomings of the conventional cloud or trusted certificate-driven techniques have motivated us to exhibit a unique Blockchain-based framework for a secure and efficient industry 4.0 system. The demonstrated framework obviates the long-established certificate authority after enhancing the consortium Blockchain that reduces the data processing delay, and increases cost-effective throughput. Nonetheless, the distributed industry 4.0 security model entails cooperative trust than depending on a single party, which in essence indulges the costs and threat of the single point of failure. Therefore, multi-signature technique of the proposed framework accomplishes the multi-party authentication, which confirms its applicability for the real-time and collaborative cyber-physical system.  

\end{abstract}


%
\IEEEpeerreviewmaketitle

\section{Introduction}

The recent US study takes us to an alarming point that our cyberinfrastructures are as vulnerable as we are to the COVID-19. Internet Crime Complaint Center (IC3) of the Federal Bureau of Investigation (FBI) reports that there has been a 300\% increase in cybercrime since the beginning of the latest pandemic. They further warn that 95\% of the recorded breaches targeted mostly three critical infrastructures such as government, technology, and retail industry.  While investigating the threat type, Cisco unveiled Advanced Persistent Threat (APT), ransomware such as Zero-day, Spyware, and Botnet Virus, occupy more than half of the total attacks with a loss of more than \$500k. Besides, it is worth frightfully mentioning that about 6 trillions of budgets are globally projected to tackle the potential crisis, which was only 1 trillion in 2018. Since the last decade, this budget has been overgrowing as the world has experienced the latest industrial  iteration namely industry 4.0\cite{i4}. This critical infrastructure has laid the foundation for the desired smart systems, where the convergence happens between machines and humans, depending on data.  Sensor-generated data play a vital role in monitoring the manufacturing process, such as predicting maintenance, detecting the equipment anomalies, etc. For example, the data-driven predictive maintenance systems supervise critical tasks such as electrical insulation, vibration or temperature analytic, etc. As a critical component, if data fails to comply with the security standard, all the actions associated with data would undeniably counterfeit the entire industrial ecosystem.
Even after deploying all the advanced technologies, i.e., Cisco Forrester Zero Trust security (2019) or Software Intel Guard Extension (SGX-2015) the world of the state-of-the-art has encountered unprecedented and, in some cases, unknown (i.e., WeLeakData, January 2020) cyberattacks.  The latest attacks, including Advanced Persistent Threat (APT), Drive-by, Spear phishing, Session hijacking, Zero-day exploit, were triggered off on account of inadequate and untrusted data and network protection. Existing industrial security solutions inlcuding some public Blockchain-based approach (i.e, CertLedger) are designed after depending on the trust provided by the trusted third party (TTP) such as PKI (public key infrastructure) or cloud-driven trusted certificate authority. The system with such remedies has experienced adversaries along with several other issues such as adaption, delay, expenses, latency, etc.  Moreover, any system taking service from a single trusted entity holds an imminent threat of data breaching or the intimidation of single point of failure (SPoF). 

Therefore, the paper motivates demonstrating a framework to solve the existing security challenges (i.e., multi-party consents, the centralized trust of trusted certificate authority, etc.) for the Industry 4.0 Cyber-physical System (CPS). The blockchain-based Multi-Signature (MS) mechanism ensures collaborative trust rather than the traditional PKI-like single signing technique that often adheres to a single-point trust dependency and transparency loophole.

\subsection{Contributions and  Organization}
Industry 4.0 CPS need to incorporate
collaborative trust building
rather than delegating the trust authority to a single honest or semi-honest  entity\cite{i4}. Therefore, the demonstrated Blockhain based framework proposed to ensure security by a special certificateless authentication technique developed upon the MS scheme.  The framework is reluctant to present-day cyber attacks. For precise clarification, the specific contributions can be briefed as below-

\begin{itemize}
    \item \textbf{Certificateless multi-party security:}
    The framework utilizes smart-edge computation, works over peer-to-peer (P2P) consortium, and eliminates the centralized trust model, i.e.,   dependency on the single PKI certificate authority (CA) for public keys. In addition, multiple Industry stakeholders collaboratively authenticate both the device identity and data. This unique feature guards data against forgery, enhances trust and overcomes the single-point-of-failure (SPoF).
    \item \textbf{Efficient and reliable network}: Consortium Blockchain peers of the proposed framework validate and track device registration and communication activities into a transparent ledger inside the restricted channel. In consortium consensus protocol, the leader proposes the next block that significantly lower the reward costs. It shields against failures by subduing the influence of the potential malicious peers and finalizes the agreement on a new transaction without multiple confirmations, thus no waiting period requires after a particular block addition. Besides, the permissioned node selection (i.e.,PBFT, Chaincode, as in the Hyperledger Fabric (HLF) or Iroha (HLI), to be discussed in section III) and the banning of malicious node keeps the industry 4.0 cyber-space secure and threat free\cite{dht}.
    \item \textbf{Distributed off-chain storage:} Instead of conventional cloud or database, the proposed framework adapts storing data at the distributed hash table (DHT), i.e. InterPlanatry File System (IPFS) or Kademila. The salient feature of P2P, version-controlled and content-addressed file system, ensures faster data transfer and reduces server dependency to save additional spending and bandwidth\cite{dht}.
    
\end{itemize}
  
   The proposed technique obtains collaborative trust from Consortium Blockchain (CBC) peers instead of conventional PKI-driven CA, i.e., VeriSign, DigiCert, etc. Thus, it can circumvent the chance of SPoF on account of the TTP's betrayal. BC consensus in line with the multi-party consents (through multi-signature), ensure the entrusted security of several industrial stakeholders. Simultaneously, it reduces both response latency and risk associated with the trusted certificates issued by a single entity, i.e., Membership Service Provider (MSP), VeriSign, etc.

   The rest of the article is organized as follows- the next section illustrates the suitability of the Consortium Blockchain technology, the security requirements and the typical adversaries of the critical Industry 4.0 enterprise. Then the subsequent section illustrates the proposed BC-based certificateless framework and its working components. Before the conclusion, the framework deployment and evaluation section reinforces the claims of improving the performance configured on the CBC, namely HLF. 


\begin{figure}[htb!]
    \centering
    \includegraphics[width=\linewidth]{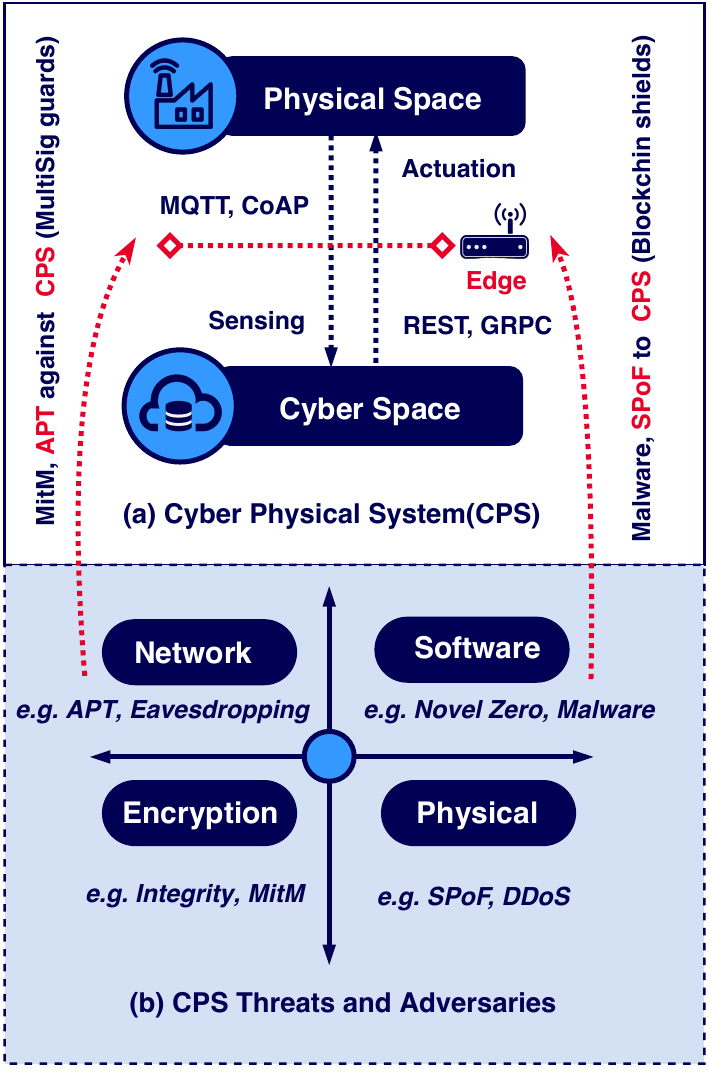}
    \caption{  a) Communication flow of CPS b) CPS attacks BC can guard against}
    \label{fig:fig_1}
\end{figure}


\begin{figure*}[htb!] 
\centering\includegraphics[width=\linewidth]{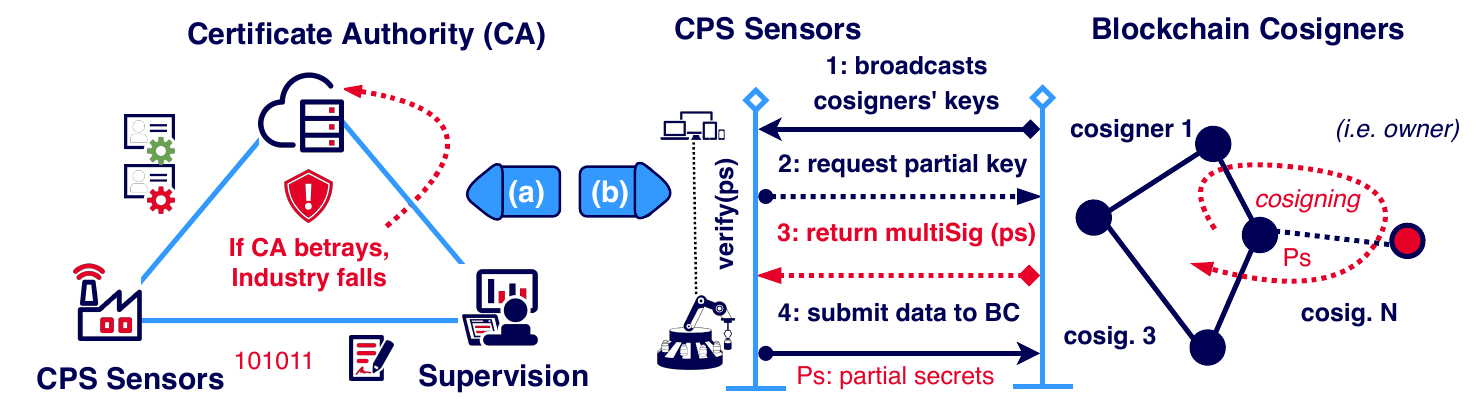}
\caption{(a) Industry 4.0 CPS Trust and failure Challenges with Certificate Authority (b) Multisignature and BC Solution not requiring the Certificate Authority}
 \label{fig:fig_2}
\end{figure*}

\section{Blockchain for Critical Industry 4.0  CPS}
Any trusted and reliable critical system essentially requires to be consistently capable of enduring prohibitive loss all the way it evolves. From the conservative development perspective, innovating such a decisive and crucial system naturally demands proven techniques rather than the naive approaches that often appear appealing at its first sight. Though, BC technology initially had attracted inundated public attention but afterward established its impressive attributes to unravel the advanced security breaches, i.e., APT, Zero-day attack, etc. as shown by Figure \ref{fig:fig_1}. It also helps explaining that CPS refers to the firm conjoining and coordination between cyber and physical resources. The CPS comprises software and physical components where each component works on the different temporal and spatial levels and continuously interacts with one another. BC with MS seems to be a promising alternative in response to those advanced attacks as shown by Figure \ref{fig:fig_1} (b) respectively.


\subsection{Critical Industry 4.0 Issues with Public Blockchain}
BC is resistant to modification as if once data recorded, the block cannot be altered \cite{bciot}. The link between subsequent blocks breaks and demands consensus of the participating nodes if any changes occur in the previously committed blocks. Unlike the PKI or cloud-driven CA, a BC requires running consensus i.e., proof of work/stake (PoW/PoS) and associated smart contracts (SC) protocols before updating a new block to the shared ledger. The reward and energy-intensive and mining nodes driven consensus works for public BC which best suits where an utterly untrusted network is required to be safe\cite{privacy}. It is unparalleled in cryptocurrency \cite{btc} but logically unsuited for the critical industry, i.e., smart-grid,  food or health system, road signaling, etc. due to its slower performance.

\subsubsection*{Consortium Blockchain Advantages}

 Whereas CBC such as Hyperledger or  Quorum, has a selective setup where only trusted and invited members are allowed to join the network\cite{btc}. As new block verification does not require running PoW-like consensus, CBC can obviate additional expenses for setting up energy-intensive miner-nodes and compensate needed rewards or incentives. Instead, it adapts the fault tolerance consensus mechanism, where the leader proposes the next block to lower reward costs\cite{con}. It protects system from failures by subduing the influence of the potential malicious peers and thus finalizes the agreement on a new transaction without multiple confirmations. The transaction processing rate of this permissioned nature blockchain\cite{dht} is significantly higher than that of the miner-driven public BC,(i.e., for Bitcoin and Ethereum, about 4 to 5 transaction and for HLF, about 3,000 to 20,000 transaction per second) making it a convincing choice for the proposed data protection framework of the Industry 4.0 CPS.


\subsection{Challenges of the Cyber-physical System}

Since the 2011 Hannover Fair by German government, the world has experienced the latest iteration of the industrial ecosystem namely industry 4.0 because of its unique integration of the CPS\cite{i4}. The ultimate goal of building such automated connections between cyber space and physical space range from enhancing productivity to boosting revenue where BC comes to plug-in-play after cooperating with its existing cybernet technologies such as big data, deep Learning, etc \cite{Blockchain}. The critical industry 4.0 infrastructure demands security assurance from multiple stakeholders, instead of single trusted party such as CA. Figure \ref{fig:fig_2} (a) illustrates how the entire industry 4.0 ecosystem falls due to CA's betrayal \cite{li2018blockchain}.

\section{Blockchain based Security Framework}
\begin{figure*}[htb!] 
\centering\includegraphics[width=\linewidth]{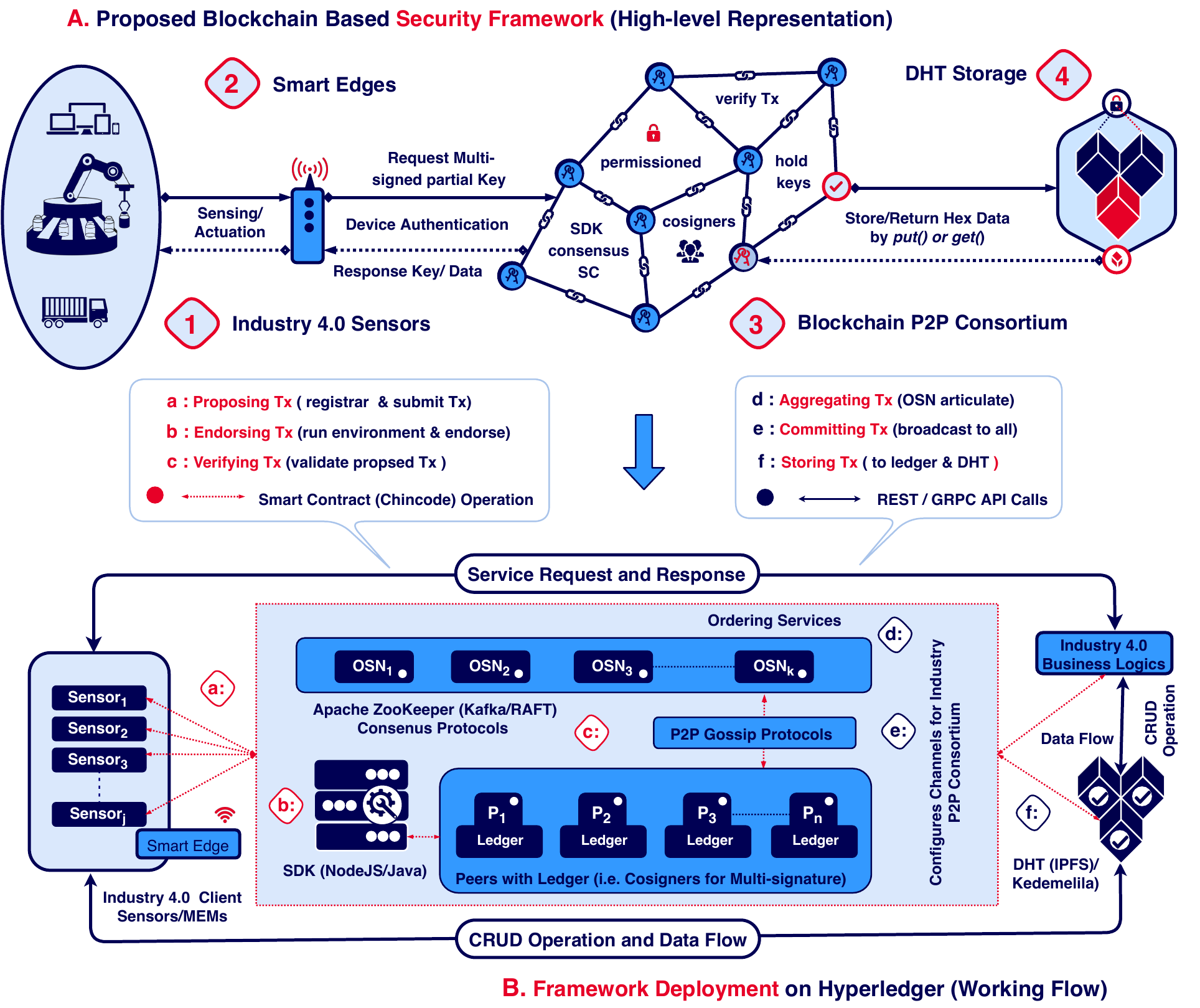}
\caption{A. High-level Representation of the Proposed Security Framework, B. Working Flows as per the Framework Configuration}
 \label{fig:fig_3}
\end{figure*}

The proposed security framework for critical industry 4.0 CPS works on four different levels, where sensor and smart edge devices work in the physical space, and consortium Blockchain network or DHT work in the cyberspace as demonstrated by Figure \ref{fig:fig_3} (A). Based on the 4 levels of communication, the working principle can be simplified through 3 different core stages. Firstly, the machine sensors get registered and establish communication using a BC-based MS technique. Secondly, the registered sensors submit data to the BC network subjective to be verified by BC consortium. Thirdly, the data is stored in the DHT after recording the log into the BC ledger\cite{dht}. As claimed in the earlier subsection, the framework utilizes edge computation instead of cloud computing\cite{edge}. The reasons are listed as following before explaining the steps within the proposed framework.


\begin{itemize}
    \item  As usually sensors have a weak computational ability that makes them prone to failure while interacting with its heavy storage node such as cloud \cite{Blockchain}. BC can quickly adopt smart edge computations with neater storage solution.
    \item As the framework requires signing, smart edge device can securely do the required solving cryptography that the light-weight sensors are unable to do. 
    \item  Besides, it can send data to the P2P DHT storage and can easily identify the DHT-address \cite{edge}. 
\end{itemize}

 In ordinary certificate-less authenticated encryption (CLAE) or identity-based encryption (IBE), partial key intends to resolve the key-escrow problem (keys are essential to decrypt). The proposed technique replaces it PKI-like key generation centre (KGC, works as a trusted center similar to CA) by adopting a CBC consisting of collaborating peers of the Industry 4.0 CPS. In the proposed framework, the BC peers, using the underlying system parameters, first generate and then dissipate the key pairs where the private parameters are known as the partial key. The partial secret (PS) of the BC consortium and the sensor's random secret in conjunction constitute the desired private key. It secures data from the physical space, i.e., embedded machine sensors to cyberspace, i.e., BC or DHT storage, purposing to check out single-point-dependency and guarantee mutual collaboration\cite{li2018blockchain}.

\subsection{Certificateless Sensor Authentication by Multisignature}
MS means applying more than a single key in authentication, which seems so promising for multi-party industry 4.0 security. For example, in an industry 4.0 setup, the sensitive data generated by expensive equipment need to be equally consented by its owner, operator, insurer, or buyer. Each of the parties could be geographically dislocated but cooperate for a single mechanical event. In such a scenario, trusting a single party, i.e., CA, is expensive and not consistently reliable. The principal objective of signing data is to ensure that data has come from an authentic source, and upon producing the same message-digest to the recipient, it further proofs data integrity.  Therefore, the framework incorporates MS to increase the trust and pertaining reliability and handle prohibitive costs all the way the industry 4.0 CPS evolves\cite{multiparty}.

\subsubsection*{Blockchain Replaces the Certficate Auhority}

As illustrated by Figure \ref{fig:fig_2} (b), the BC peers ensure generating a partial secret  before responding to the sensor devices. The process begins through broadcasting the multiple public keys among the industry 4.0 sensors by BC consortium. The employed signature is a variant of the existing  Boneh–Lynn–Shacham (BLS) signature scheme, which works upon two different elliptic curves (EC) over a finite prime field\cite{bls}. It has implemented a similar partial secret concept of IBE on top of BC  and thus eliminates the Key $KGC$ \cite{li2018blockchain}. Suppose there are several industry stakeholders who have been collaborating with their public-private key pairs. Two different hash algorithms produce respective message digest throughout the process. A sensor runs system's security parameters to bring out the secret values. Then it sends their identity (ID) and after encrypting it using the aggregated public keys. BC peers verify and generates the PS and cosign by all of its participating peers. The framework initially requires signing by all; however, it is not necessarily required as its supports threshold signing implicitly depicted by the curved arrow of Figure \ref{fig:fig_2} (b). The industry sensors then verify and decrypt the PS and accomplish their public-private keys to send data to the BC. The proposed framework excludes PKI-like certificates by implementing the intermediary PS which in essence guarantee desired trust to the sensors\cite{multisig}. 

\subsection{Transaction Verification and Storing}
 Once industry 4.0 sensors are successfully registered to the BC consortium, they are ready to submit data as transaction proposal, as illustrated by Figure \ref{fig:fig_3} (B). Data  usually gets transaction (Tx) fashion that includes the \textit{identity}, \textit{timestamp}, and \textit{action} of the CPS sensors. \textit{Action} may vary as per the request type, i.e., \textit{update}, \textit{store}, or \textit{access} to DHT. BC peers have to meet two necessary conditions to verify a transaction:
\begin{itemize}
    \item [] \textit{i)} The public key obtained associates with the identity?
    \item [] \textit{ii)} Can the transaction received be verified?
\end{itemize}
 The transaction verification includes the steps illustrated by the framework configuration on HLF CBC \cite{GDPR2019}. Figure \ref{fig:fig_3} (A) demonstrate the proposed figure with 4 levels (1 to 4) and the extended Figure \ref{fig:fig_3} (B) illustrates the following steps from \textit{a} to \textit{f}.

\textit{\textbf{a) Propose}}: Client CPS devices initiate the process by registering the devices to the BC, as explained in the earlier section. Then it starts constructing the encrypted transaction proposal using the private keys  and invoke the SC with the help of Software Development Kit (SDK) and required application programming interfaces (APIs).
    
\textit{\textbf{b) Endorse}}: SDK requests for endorsement, and BC peer verifies the Tx after authenticating the device identity (id) subject to ensure that data is coming from the right sources.
    
\textit{\textbf{c) Verify}}: The endorsement verification requires meeting the policy, i.e., business logic agreed by all stakeholder peers, by running the respective SC operation. The SC takes a Tx proposal as input and returns a multi-signed \textit{YES}  or \textit{NO}  in response to the SDK apps against step \textit{b}).
If the Tx proposal is determined as query function, the SDK App returns the data upon query execution with the help of respective APIs (i.e OAuth 2.0 REST API). In either case, the SDK apps forward the transaction with the required operations such as C-create, R-retrieve, U-update, and D-delete with the endorsement. 
    
\textit{\textbf{d) Aggregate}}: After receiving the verified consent, the SDK apps aggregates all consents into single (or multiple if required) Tx and disseminates those to the Ordering Service Nodes (OSNs). The OSN works on the consensus protocols such as crash fault tolerance (CFT) with RAFT (etcd.io libraries) or practical byzantine fault tolerance (PBFT) within Apache Kafka platform.
    
\textit{\textbf{e) Commit}}: The Tx then relayed to the OSN, all channel peers validate each Tx of the block by specific SC validation and checking through concurrency control version. Any Tx fails the process is marked with an invalid flag inside the block. Thus, a desired new block is committed to the ledger.
    
\textit{\textbf{f) Store}}: The Gossip data dissemination protocol of the OSN concurrently broadcasts ledger data across the consortium to ensure synchronization of the latest version of the ledger.  The pointer of the data is set out in the BC ledger, and at the same time, data-address is securely stored on the offline or online DHT data repository upon IPFS or HL CouchDB. It confirms the unique benefits such as traceability, P2P adaptability etc. that are already outlined in the earlier contribution section. DHT incorporation makes the framework robust, self-organizing, scalable, and fault-tolerant against  routing attacks and false query injection attacks. The proposed framework mostly suits the DHT protocol, however, we have initially integrated IPFS and Kademlia considering the fixed-sized routing and malware attacks.

\begin{figure*}[htb!] 
\centering\includegraphics[width=\linewidth]{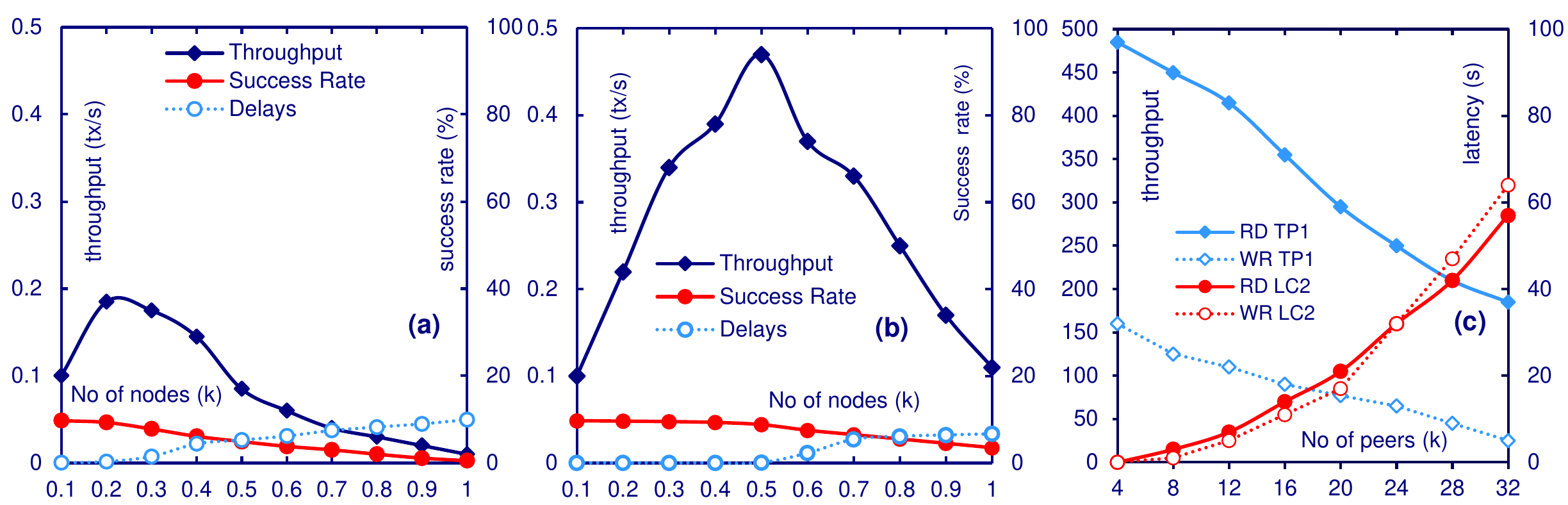}
\caption{READ (a) vs WRITE (b) performance and Throughput (TP) Latency (LC) for two different setup, with 300 Workload, another with 500 workload}
 \label{fig:fig_4}
\end{figure*}

\subsection{Smart Contract and Consensus Implementation}
As a transaction protocol or program, smart contract resiliently executes, manages assigned events and actions as per the agreement often called \textit{contract}. Because of its self-governing features, it can avoid the need of a trusted intermediary, prohibitive costs and accidental losses\cite{sc}. The framework was deployed both on IBM HLF and HLI, which facilitates writing customized Smart contract called the chaincode (CC) for specific tasks. Sample CC are available in the referred project link \footnote{Github Project Link:https://github.com/rahmanziaur/IIoTConsortiumBC}, provides the founding environment for authentication, access control, and validation. Being a platform-independent, HLF supports any language to write its codes\cite{GDPR2019}; however, because of related online resources, we mostly preferred \textit{Go} and, in some test-cases \textit{Java}. The customized CC replaces the existing codes for Membership Service and its dependencies. Consortium Blockchain entails default Membership Service Provider (MSP) that works as PKI-like Certificate Authority (CA). It provides ledger/other CC accessing APIs, state variables, or transaction context. 
The policy is defined as an access control list (ACL) for simplicity. However, as explained earlier in the BC suitability subsection, BFT consensus has salient advantages over the existing PoW driven approaches\cite{dht}. HLF incorporates Apache Kafka that eases the execution of consensus, such as BFT, CFT, considering enterprise security requirements. CFT can withstand even half of the total nodes fail but do not guarantee adversary nodes what BFT does. It retains working even one third of total nodes falls. PBFT protocol seems more resilient to the system failure; however, the proposed framework is reluctant to both BFT and CFT\cite{bciot}. 


\subsection{Framework Threat and Trust Model}

 The communication between edge devices \cite{edge} (i.e. Dell, Hawaii, AD-Link gateway), accessories, and critical industry 4.0 machine sensors (i.e. motor vibration, force or humidity sensors) are secure. Besides, consortium peers (i.e. endorsement, anchor peers of the HLF/HLI  are assumed to be trusted, semi-honest or honest-but-curious who issues multisigned PS instead of traditional certificates\cite{GDPR2019}. However, MS and its associated encryption technique together ensure secure communication between edge and BC level and guard network and data against data breaching, malware injection, etc. 

\section{Blockchain Deployment and Evaluation}
The CBC of the proposed framework was deployed inside the Caliper evaluation toolkit for Hyperledger, which helps measuring a particular BC deployment with a set of previously defined enterprise use cases. IBM admits that no general tool provides performance evaluation for BC while releasing Caliper\cite{GDPR2019}. The extensive use-cases were set to overlap the industry 4.0 client requirement that generates the benchmark data. However, the latest version of the nodeJS package manager (NPM 8.0.1), docker, and  curl  were installed to setup the run-time environment inside the Ubuntu 18.04 LTS with 32 GB of memory where \textit{python2} , \textit{make}, \textit{g++} and git ensure additional supports. A typical configuration for the BC evaluation should have programs called \textit{Test Harness} that include client generation and observation and the deployed BC System Under Test (\textit{SUT}) and the RESTful SDK interfaces\cite{GDPR2019}. There are four (04) fundamental steps required to finalize the evaluation process as following. 

\begin{itemize}
    \item [] \textit{a}) A local \textit{verdaccio-server} for package publishing
    \item [] \textit{b}) Connecting the repository to the server 
    \item [] \textit{c}) Binding the \textit{CLI} from the server-side, and
    \item [] \textit{d}) Running the  benchmark as per the configuration file
\end{itemize}

 Then the performance benchmark with the tasks, firstly, invoke policy checking functions to READ and WRITE Tx into the ledger. Secondly, setup multiple test-cases for 4 to 32 number of peers representing industry 4.0 stakeholders. Finally, allocating the workloads from 100  to 1000 transaction per second among the peers representing the CPS sensor device data population. The throughput (TP) and latency (LC) are calculated based on the READ (RD) and WRITE (WR) performance where the delay LC is the time between the time of request and response of the transaction. TP is the ratio of committed Tx at a particular time.

\subsection{Performance Analysis}
 The Hyperledger Caliper benchmark results show the performance based on four measurement metrics \textit{success rate}, (transaction successfully committed against the submitted proposal), \textit{latency} and the \textit{throughput}, and the \textit{resource consumption} for different test cases. The Figure \ref{fig:fig_4} (a) shows the system performance under the different number of sensor workload ranging from 100 (0.1k) to 1k workload where the HLF network occupies two (02) CC, four (04) peer nodes and three (03) OSNs running on Apache Kafka (or ZooKeeper) for PBFT. As seen in the Figure \ref{fig:fig_4} (b), the WRITE has 185 at 0.2k workload with the maximum success rate of 93\% and an average delay of 5 seconds. Notwithstanding, operation seems to have higher throughput (up to 470 in maximum) on a similar success rate at its best. The average delay appears to be half of the WRITE's delay, as writing has to incorporate OSNs on Apache Kafka. The benchmark evaluation explicitly illustrates that the setup configured has lower performance for higher number sensor-workload through the theoretical solution proves the consortium BC has significant adaptability for a higher number of nodes. As investigated, the local workload processing bottleneck affects throughput and latency. HLF Tx flow works demands enough responses against the submitted transaction proposals, in case the responses are queued due to network overhead, bandwidth, or processing-load consequences the latency-raising. On top of that, the general-purpose workstation configuration slower the evaluation for higher workloads. Here, the result portrays the relation between performance and scalability based on the previously executed READ, WRITE Operations. OSN and peer configuration left resembling initial setup with two test-cases run for 300 and 500 workloads. As depicted by the Figure \ref{fig:fig_4} (c), the deployed framework has lower scalability. For the first test-case (300 workloads), the throughput and latency reach 150 transaction per second and 64 transaction per second. However, for the other test-case, it comes with lower throughput and higher latency concerning the number of nodes ranging from 4 to 32. Caliper toolkit allows to run the node subset that endorses a particular SC. The proposed framework upon HLF Caliper benchmark fetches higher latency (i.e., about a minute for 32 peers) due to computation constraints. Besides improving the consensus in terms of a suitable MS scheme, further scopes include applying an enterprise standard ( i.e.,  512 GB of RAM or more) system to improve the latency overhead currently it suffers from. 

 For the same setup, the framework's average response time is several times lesser than the system with the MSP. By default, MSP provides certificate service, i.e.,  Hyperledger Fabric MSP, VeriSign, etc. inside the consortium Blockchain network \cite{GDPR2019}. The result obtained shows that the proposed framework without MSP or any other certificate provider, including the trusted KGC, responds within 1 to 16 milliseconds. However, it delays 40 to 242 milliseconds with the default MSP of the Hyperledger CBC deployment. The response latency varies with the increase of workloads (number transactions to process)  and co-signing BC peers. The proposed MS aligned framework (no MSP as discussed earlier) absorbs almost 4-times lower power than the public BC with traditional signing techniques. Also, it spurs about 50 percent lesser CPU than the consortium BC lined with the conventional MSP. Besides, the certificate-less mechanism, special remote procedure call, namely gRPC and mining-free consensus protocols, i.e., PBFT of HLF, jointly economize the energy and CPU usage it needs.

\section{Conclusion and Work to Ahead}
Relying on a single trusted party challenges the industry trust, makes it exposed to several advanced cyberattacks, and causes single-point-of-failures. Admittedly, critical industry infrastructure demands to incorporate cooperative trust-building rather than trusting a single entity. In response to such security issues, we have demonstrated the BC-based framework that ensures security without existing PKI-like certificates. The contribution includes unique adaption of multi-signature inside a reliable BC consortium subject to data protection with an efficient storing technique. The achieved performance  support the framework applicability for the critical industry 4.0 enterprise CPS. 

\ifCLASSOPTIONcaptionsoff
  \newpage
\fi

\bibliography{sample}
\bibliographystyle{ieeetr}

\begin{IEEEbiographynophoto}{Ziaur Rahman}
 is currently a PhD scholar at RMIT University. He is a graduate member of ACM, Australian Computer Society and IEEE. His research interest includes Blockchain, IoT etc.
\newline
\newline
\textbf{Ibrahim Khalil}
is an Associate Professor at the RMIT University and previously worked in Silicon Valley, USA. He was the recipient of IEEE LCN Best Paper award and the Fritz-Kutter award. He has served as PC member of several IEEE conferences. His research interests overlap Blockchain, IoT etc.
\newline
\newline
\textbf{Xun Yi} is currently a Professor at the RMIT University. He has led several projects and been an expert member of Australian Research Council. He is an Associate Editor of the IEEE TDSC, and PC member of over 40 international conferences. His research interests include Blockchain, Cyber Security, etc.
\newline
\newline
\textbf{Mohammed Atiquzzaman} is a Edith Kinney Gaylord Presidential professor at the University of Oklahoma. His researches are supported by NSF, NASA and the US Air Force. He serves as the Editor-in-Chief of Journal of Network and Computer Applications, Co-Editor-in-Chief of the Computer Communications Journal and Editor of IEEE Transactions on Mobile Computing, IEEE JSAC, etc.
\end{IEEEbiographynophoto}

\end{document}